\documentclass[11pt]{article}

\usepackage{acl}

\usepackage{times}
\usepackage{latexsym}

\usepackage{comment}
\newcommand{\ie}{\textit{i}.\textit{e}.}

\usepackage{amsmath,amssymb}

\usepackage[T1]{fontenc}

\usepackage[utf8]{inputenc}

\usepackage{microtype}
\usepackage{placeins}
\usepackage{inconsolata}

\usepackage{graphicx}
\usepackage{booktabs}
\usepackage{multirow}
\usepackage[table]{xcolor}

\title{Decoding Strategies for Diffusion-Based ASR:\\A Systematic Evaluation of Confidence-Based Thresholding}

\author{Jeong Hun Yeo$^1$ \quad Minsu Kim$^{2,*}$ \quad Hyeongseop Rha$^1$ \quad Yong Man Ro$^{1,\dagger}$
\\$^1$ KAIST, Daejeon, Republic of Korea \\
    $^2$ Google DeepMind, Tokyo, Japan \\
\small{\texttt{sedne246@kaist.ac.kr, ms.k@ieee.org, ryool\_1832@kaist.ac.kr, ymro@kaist.ac.kr}}}

\begin{document}
\maketitle
\def\thefootnote{}\footnotetext{$^*$Work done in an advisory role only. $^\dagger$Corresponding author.}

\begin{abstract}
While LLM-based Automatic Speech Recognition (ASR) achieves high accuracy, its speed is limited by sequential autoregressive decoding. Diffusion Language Models (DLMs) offer a parallel alternative, yet their decoding strategies remain under-explored in ASR contexts. This paper analyzes three decoding schemes for DLM-based ASR: fixed-number, static confidence threshold, and dynamic confidence threshold. We introduce a round-wise analysis of decoding progress using Negative Log-Likelihood-based uncertainty as a proxy for prediction reliability. Our results show that both threshold-based strategies provide a better accuracy–speed trade-off than fixed-number schemes. This behavior is associated with the more concentrated confidence distribution observed in the evaluated ASR settings: many tokens reach high confidence early, enabling multiple tokens to be committed in early decoding rounds while lower-confidence tokens are deferred to later rounds. The static-threshold strategy achieves accuracy close to autoregressive decoding at lower decoding cost.
\end{abstract}

\section{Introduction}
Automatic Speech Recognition (ASR) has advanced through self-supervised pretraining~\cite{baevski2020wav2vec, schneider2019wav2vec, hsu2021hubert, chen2022wavlm}, large-scale weakly supervised training~\cite{radford2023robust, pratap2024scaling, peng2024owsm}, and LLM-based decoder modeling~\cite{grattafiori2024llama, bai2023qwen, chen2024salm, wu2023decoder, chu2023qwen}. However, inference in LLM-based ASR remains computationally expensive, primarily because autoregressive (AR) decoding generates tokens sequentially, requiring repeated forward passes through large decoder models.

Before the adoption of LLMs, non-autoregressive ASR methods such as Imputer~\cite{chan2020imputer}, Mask-CTC~\cite{higuchi20b}, and Align-Refine~\cite{chi2021align} had already reduced reliance on sequential decoding through iterative parallel prediction and refinement of token sequences and latent alignments. This parallel-refinement paradigm has since been revisited in ASR~\cite{wang2025audio, tian2026dllm} through the adoption of Diffusion Language Models (DLMs)~\cite{nie2025large, ye2025dream}. These models leverage linguistic knowledge acquired through large-scale language pretraining and enable bidirectional parallel decoding by iteratively predicting multiple masked positions and committing high-confidence tokens.

Since token commitment in DLMs relies on heuristic confidence criteria, prior work has explored fixed-number, static-threshold, and dynamic-threshold decoding strategies~\cite{fu2025bits, ben2026accelerated, nie2025large, chang2022maskgit, yu2025dimple, wu2025fast}. Existing DLM-based ASR systems, however, employ individual strategies in isolation: Whisper-LLaDA~\cite{wang2025audio} uses fixed-number decoding, whereas dLLM-ASR~\cite{tian2026dllm} uses static thresholding. Neither systematically compares alternative commitment strategies nor analyzes why their behavior differs. We therefore compare the three strategies in terms of accuracy, decoding speed, and round-wise progress.

Concretely, we analyze decoding progress through prediction reliability and token throughput. Since insertion and deletion errors break alignment across decoding rounds, round-wise accuracy cannot be measured directly. We therefore use cumulative token-level Negative Log-Likelihood (NLL) as a proxy for prediction reliability, with an AR configuration of the same DLM as a reference. We measure throughput by counting newly unmasked tokens per round and examine confidence distributions to characterize differences in commitment behavior.

Our main findings are threefold. First, static thresholding offers a favorable overall accuracy--speed trade-off across test sets and two DLM backbones, achieving comparable WER to dynamic thresholding with lower RTF, while both threshold-based strategies outperform fixed-number decoding in the evaluated settings. Second, round-wise analyses of NLL and token throughput indicate that the differences underlying the final efficiency and accuracy gaps emerge primarily in the early decoding rounds, where threshold-based strategies commit many high-confidence tokens. Third, this behavior is associated with confidence scores being more concentrated at high values in the evaluated ASR settings than in the examined text tasks, including math reasoning, machine translation, and summarization, which is consistent with the observed early commitment behavior.
\section{Baseline DLM-based ASR Model}
\subsection{DLM-based ASR}
Inspired by Whisper-LLaDA~\cite{wang2025audio}, we build a DLM-based ASR system that combines Whisper-medium.en~\cite{radford2023robust} as the speech encoder with LLaDA-8B-Instruct~\cite{nie2025large} as the decoder. A lightweight temporal adapter downsamples and projects the speech representations into the DLM embedding space. During training, the Whisper encoder is frozen, while the DLM is fine-tuned via LoRA~\cite{hu2022lora} and the adapter is trained jointly.

Following the diffusion formulation~\cite{nie2025large}, we randomly mask the target sequence $x_0$ with a ratio $t \sim \mathcal{U}(0,1)$ to obtain $x_t$, and train the model to reconstruct masked tokens conditioned on audio features $a$ and unmasked text context. The training objective is:
\begin{equation}
\setlength{\abovedisplayskip}{3pt}
\setlength{\belowdisplayskip}{3pt}
\mathcal{L}
= -\mathbb{E}_{t,a,x_0,x_t}
\left[
\frac{1}{t}\sum_{i=1}^{L}
m_i \log p_{\theta}(x_0^{i}\mid a,x_t)
\right],
\end{equation}
where $m_i=\mathbf{1}[x_t^{i}=\mathrm{M}]$, $L$ is the sequence length, $\mathrm{M}$ is the mask token, and $\theta$ denotes trainable parameters. We train on LibriSpeech 960h~\cite{panayotov2015librispeech} for 100,000 steps ($\approx$4.5 epochs) using 8 NVIDIA RTX A6000 GPUs, and evaluate on test-clean.

\subsection{Block-based Decoding}
The decoder operates on a fixed output canvas of $L=128$ mask tokens. We use block-based decoding, where the canvas is divided into blocks of size $B$. The blocks are processed sequentially, while diffusion unmasking is performed within each block. Fully parallel decoding corresponds to $B=L$. As in Whisper-LLaDA~\cite{wang2025audio}, once an end-of-sequence (EOS) token is committed, all remaining positions are set to EOS and subsequent blocks are skipped.
\section{Analysis Method}
\subsection{Confidence Estimation in Parallel Decoding}
In DLM-based ASR, decoding iteratively refines masked tokens. At each round $r$, for each masked position $i$ where $x_r^i=\mathrm{M}$, the model predicts a token conditioned on the audio features $a$ and the current context $x_r$:
\begin{equation}
\hat{x}_{r+1}^{i}
= \arg\max_{v \in \mathcal{V}} p_{\theta}(v \mid a, x_r),
\end{equation}
where $\mathcal{V}$ is the vocabulary. We define the confidence of this prediction as
$c^{(i)} = p_{\theta}(\hat{x}_{r+1}^{i} \mid a, x_r)$.

\begin{figure*}[t]
\centering
\centerline{\includegraphics[width=16cm]{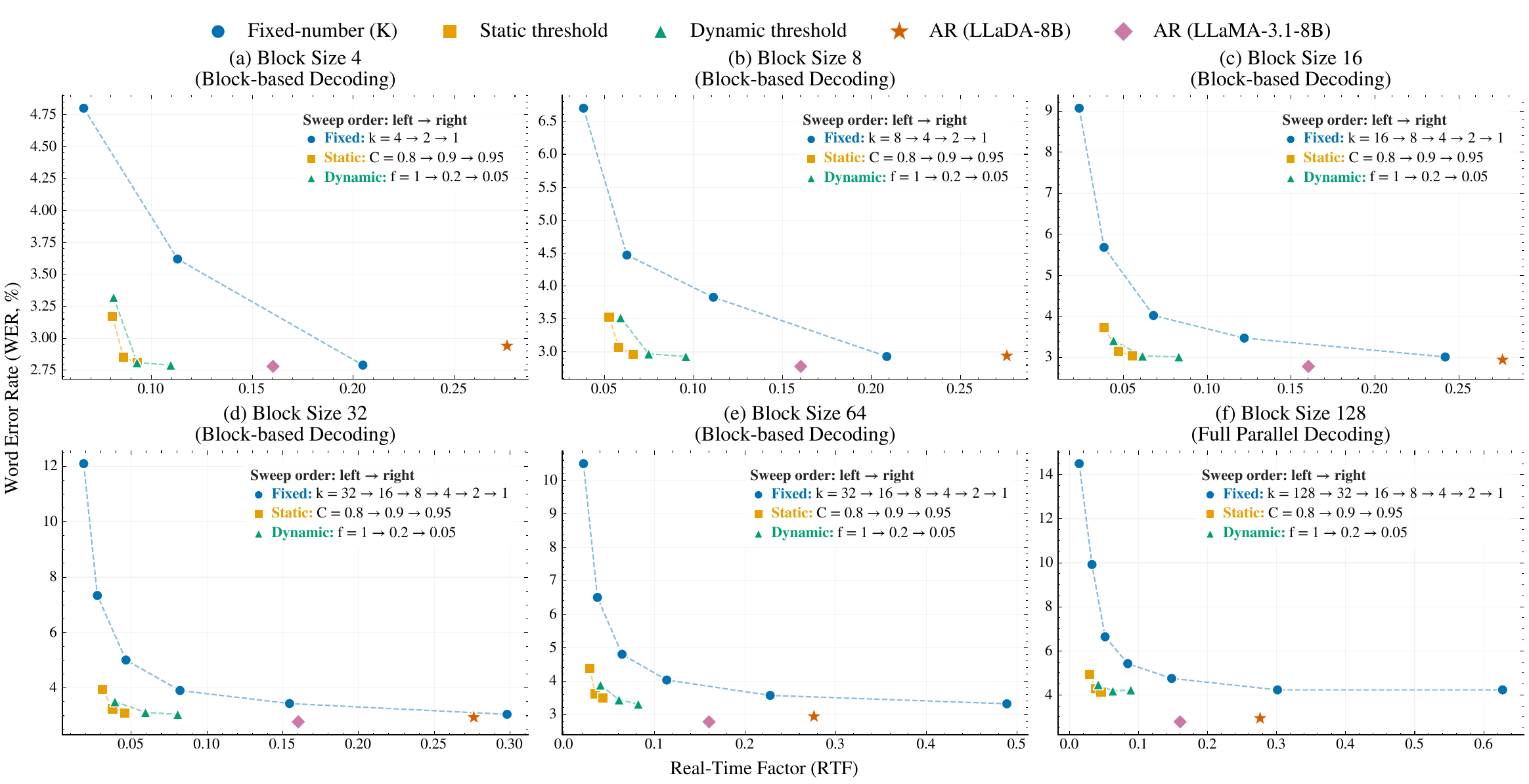}}
\vspace{-0.3cm}
\caption{WER--RTF trade-off across decoding strategies and block sizes in DLM-based ASR. Lower WER and RTF are preferred, so points closer to the origin indicate better trade-offs.}
\label{fig:1}
\vspace{-0.5cm}
\end{figure*}

\subsection{Target Decoding Strategies}
Following prior analyses of confidence-based parallel decoding~\cite{fu2025bits},
we evaluate three commonly used commitment schemes. Based on confidence scores $c^{(i)}$, we select a subset of masked positions to unmask and fix for subsequent rounds.

\textbf{Fixed-number (top-$k$) scheme.}
The model unmasks the $k$ most confident tokens among the currently masked positions in each round~\cite{nie2025large}.

\textbf{Static confidence threshold.}
The model unmasks all tokens whose confidence exceeds a threshold $C \in (0,1)$~\cite{yu2025dimple}. If none exceed $C$, the most confident token is unmasked to ensure progress.

\textbf{Dynamic confidence threshold.}
Following~\citet{wu2025fast}, confidence scores over the current $m$ masked positions are sorted as $c^{(1)} \geq \dots \geq c^{(m)}$, and the largest $k$ satisfying
\begin{equation}
(k+1)(1-c^{(k)}) < f
\end{equation}
is selected, where $f>0$ is a predefined factor. The top-$k$ tokens are then unmasked; if no $k$ satisfies the criterion, the most confident token is unmasked.

\subsection{Quantifying Progress via Token-level Uncertainty}
To analyze decoding progress, we use token-level Negative Log-Likelihood (NLL) as a proxy for uncertainty. For each predicted token $\hat{x}_{r+1}^{i}$ at round $r$, we define
\begin{equation}
U_r^i = - \log p_{\theta}(\hat{x}_{r+1}^{i} \mid a, x_r).
\end{equation}
Let $\mathcal{C}_r$ be the set of tokens committed up to round $r$. The cumulative uncertainty is
\begin{equation}
\setlength{\abovedisplayskip}{3pt}
\setlength{\belowdisplayskip}{3pt}
U^{cum}_r = \sum_{i \in \mathcal{C}_r} U_{round(i)}^{i},
\end{equation}
where $round(i)$ denotes the round in which token $i$ was fixed. We plot $U^{cum}_r$ against normalized progress $P_r=|\mathcal{C}_r|/L$.

As a reliability reference, we use an AR configuration of the same DLM, which unmasks tokens sequentially from left to right:
\begin{equation}
U^{cum}_{AR}(P)
= \sum_{i=1}^{\lfloor P L \rfloor}
-\log p_{\theta}(\hat{x}_{AR}^{i} \mid a, \hat{x}_{<i}),
\end{equation}
where $\hat{x}_{AR}^{i}$ is the predicted token at position $i$ and $\hat{x}_{<i}$ is the generated prefix. Deviations from $U^{cum}_{AR}$ serve as an indicator of the additional uncertainty associated with incomplete context in parallel decoding.
\section{Experiments}

For the main analysis metrics, we measure WER and inference speed using the Real-Time Factor (RTF), defined as the total inference time divided by the input audio duration. All latency measurements are performed on a single idle NVIDIA RTX A6000 with batch size 1, using the dual KV-cache of Fast-dLLM~\cite{wu2025fast}. As AR reference points, we use two baselines: AR (LLaDA-8B), which decodes the same DLM one token at a time from left to right, and AR (LLaMA-3.1-8B)~\cite{grattafiori2024llama}, a LoRA-tuned autoregressive decoder using the same speech encoder and adapter. Additional numerical results are provided in the Appendix.

\subsection{Decoding Strategies and Accuracy-Speed Trade-offs}
\label{sec:4.1}

We first evaluate the three diffusion-based decoding strategies on LibriSpeech test-clean by sweeping their hyperparameters: fixed-number decoding ($k \in \{1, 2, 4, 8, 16, 32, 128\}$), static thresholding ($C \in \{0.8, 0.9, 0.95\}$), and dynamic thresholding ($f \in \{1.0, 0.2, 0.05\}$), together with two AR baselines (Fig.~\ref{fig:1}). For block sizes up to 64, stricter commitment criteria (\ie, smaller $k$, higher $C$, or lower $f$) consistently reduce WER at the cost of higher RTF, reflecting the accuracy--speed trade-off in diffusion-based decoding. Compared with fixed-number decoding, threshold-based strategies achieve comparable WER with lower RTF over much of the evaluated range. Static and dynamic thresholding show similar recognition accuracy, while static thresholding often provides lower RTF, especially in Fig.~\ref{fig:1}(b,c).

Static thresholding achieves accuracy close to the AR (LLaMA-3.1-8B) baseline at lower RTF. With block size 4 and $C=0.95$, it achieves 2.81\% WER compared with the baseline WER of 2.78\%, with a $1.6\times$ speedup in terms of RTF. As the block size increases, decoding generally becomes faster and less accurate; in the fully parallel setting, static thresholding achieves 4.13\% WER with a $3.5\times$ speedup over the same AR baseline.

\begin{table}[t]
\centering
\small
\setlength{\tabcolsep}{3.5pt}
\begin{tabular}{lccc}
\toprule
\multirow{2}{*}{Test set} & Static & Dynamic & Fixed-$k$ \\
 & ($C$=0.95) & ($f$=0.2) & ($k$=2) \\
\midrule
\rowcolor{gray!15}
\multicolumn{4}{l}{\textit{In-domain}} \\
test-other  & 6.34 (.072) & 6.34 (.079) & 6.88 (.134) \\
\rowcolor{gray!15}
\multicolumn{4}{l}{\textit{Out-of-domain}} \\
TED-LIUM 3  & 10.96 (.059) & 10.97 (.066) & 11.21 (.131) \\
AMI-IHM     & 25.78 (.167) & 25.79 (.172) & 26.04 (.226) \\
\bottomrule
\end{tabular}
\caption{Cross-dataset comparison of commitment strategies. Each cell reports WER (\%) with RTF in parentheses.}
\label{tab:1}
\end{table}

\begin{table}[t]
\centering
\small
\setlength{\tabcolsep}{4pt}

\resizebox{0.98\columnwidth}{!}{%
\begin{tabular}{llccc}
\toprule
\multirow{2}{*}{Test set} & \multirow{2}{*}{Backbone}
& Static & Dynamic & Fixed-$k$ \\
& & ($C$=0.95) & ($f$=0.2) & ($k$=4) \\
\midrule

\multirow{2}{*}{test-clean}
& LLaDA-8B & 3.11 (.045) & 3.12 (.054) & 3.91 (.080) \\
& Dream-7B & 3.62 (.050) & 3.59 (.055) & 4.36 (.076) \\

\midrule

\multirow{2}{*}{test-other}
& LLaDA-8B & 6.45 (.066) & 6.43 (.075) & 7.60 (.087) \\
& Dream-7B & 6.81 (.070) & 6.80 (.075) & 7.79 (.081) \\

\bottomrule
\end{tabular}%
}

\caption{Decoder backbone comparison on LibriSpeech. Each cell reports WER
(\%) with RTF in parentheses.}
\label{tab:2}
\end{table}
\begin{figure}[t]
\centering
\centerline{\includegraphics[width=8cm]{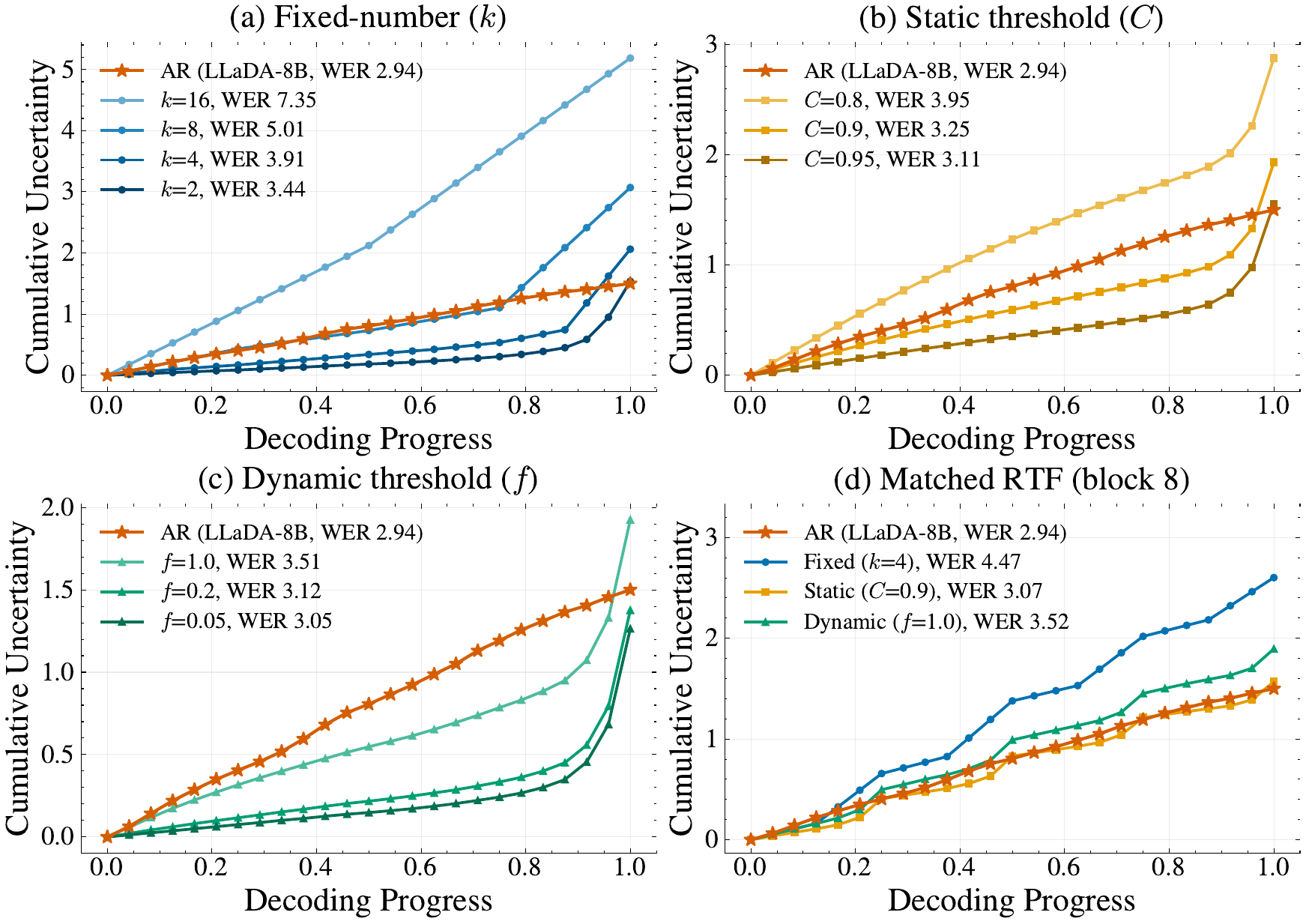}}
\vspace{-0.3cm}
\caption{Cumulative uncertainty trajectories over decoding progress. Panels (a--c) compare each strategy with the AR reference; (d) compares all strategies under matched RTF.}
\label{fig:2}
\vspace{-0.5cm}
\end{figure}

\subsection{Cross-Dataset Generalization of Decoding Strategies}

To examine whether the accuracy--speed trade-off observed on LibriSpeech test-clean extends to more challenging conditions, we further evaluate the three decoding strategies with a block size of 16 on LibriSpeech test-other~\cite{panayotov2015librispeech}, TED-LIUM 3~\cite{hernandez2018ted}, and AMI-IHM~\cite{2005ami}, covering more challenging in-domain read speech, out-of-domain prepared speech, and out-of-domain spontaneous meeting speech, respectively.

As shown in Table~\ref{tab:1}, threshold-based strategies yield lower WER and RTF than fixed-number decoding across all three test sets. At nearly matched WER, static thresholding yields approximately 3--11\% lower RTF than dynamic thresholding.

\subsection{Generalization Across DLM Backbones}
\label{sec:backbone}

To examine whether the observed decoding trends depend on the DLM backbone, we further compare LLaDA-8B~\cite{nie2025large} and Dream-7B~\cite{ye2025dream} with a block size of 32 on LibriSpeech test-clean and test-other. As shown in Table~\ref{tab:2}, similar trends are observed across both backbones: threshold-based strategies yield lower WER than fixed-number decoding, while static and dynamic thresholding achieve nearly matched WER. At comparable WER, static thresholding yields 6.7--16.7\% lower RTF than dynamic thresholding across the two backbones and test sets.

\subsection{Round-wise Progress Assessment}

For this analysis, we normalize progress over the first 32 text-token positions and consider utterances with at least 32 text tokens, setting $P_r=1$ when all 32 positions are fully unmasked.

\subsubsection{Prediction Reliability using Uncertainty Trajectory}

Measuring round-wise WER is impractical because insertion and deletion errors break alignment across decoding rounds. We therefore use cumulative uncertainty as a proxy for prediction reliability. As shown in Fig.~\ref{fig:2}(a-c), a larger final gap from the AR reference generally corresponds to higher WER, consistent with an association between cumulative uncertainty and recognition error. Most trajectories steepen near the end, suggesting that later rounds commit more low-confidence tokens; early deviation above the AR trajectory may reflect higher uncertainty in early commitments.

Under matched RTF (Fig.~\ref{fig:2}(d)), static-threshold decoding tracks the AR trajectory most closely and achieves the lowest WER among the compared configurations: 3.07\% versus 4.47\% for fixed-number and 3.52\% for dynamic thresholding.

\begin{figure}[t]
\centering
\centerline{\includegraphics[width=8cm]{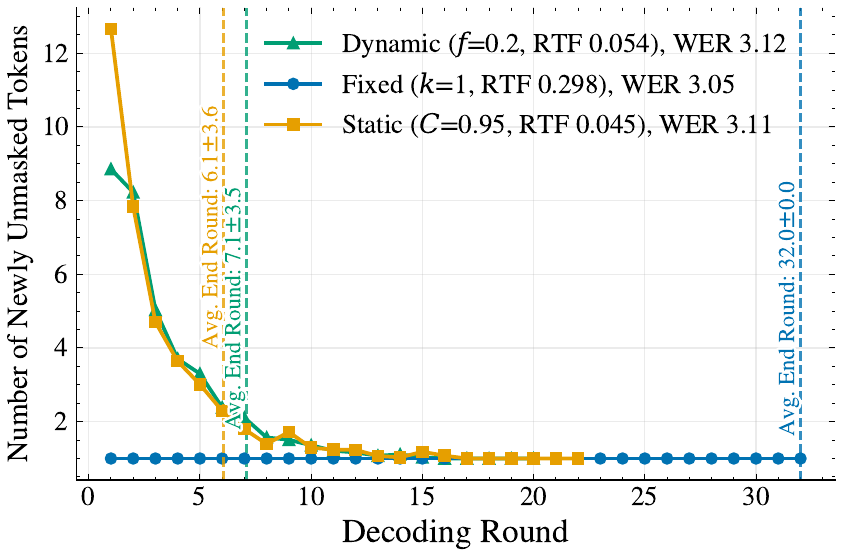}}
\vspace{-0.3cm}
\caption{Per-round token throughput under comparable-WER settings. Dashed lines indicate mean stopping rounds.}
\vspace{-0.5cm}
\label{fig:3}
\end{figure}

\subsubsection{Inference Speed using Number of Unmasked Tokens}

To examine where speed differences occur, we compare per-round token throughput under comparable WERs (3.05\%--3.12\%): static thresholding ($C=0.95$), dynamic thresholding ($f=0.2$), and fixed-number decoding ($k=1$, included as a comparable-WER operating point). As shown in Fig.~\ref{fig:3}, static thresholding finishes in the fewest rounds on average (6.1), followed by dynamic thresholding (7.1) and fixed-number decoding (32.0), consistent with their RTFs of 0.045, 0.054, and 0.298, respectively.

The observed difference is concentrated in the early decoding rounds. Static thresholding exhibits a heavy-start pattern, committing many tokens early before tapering to final refinement, whereas dynamic thresholding proceeds more steadily and fixed-number decoding is constrained by its per-round budget.

\subsection{Confidence Distribution and Decoding Efficiency}
\label{sec:confidence_dist}

To examine the confidence distribution associated with the early commitment behavior, we compare commit-time confidence across ASR and several text tasks by committing the highest-confidence token at each step. The text tasks are evaluated using the base LLaDA-8B-Instruct model.

\begin{figure}[t]
\centering
\centerline{\includegraphics[width=8cm]{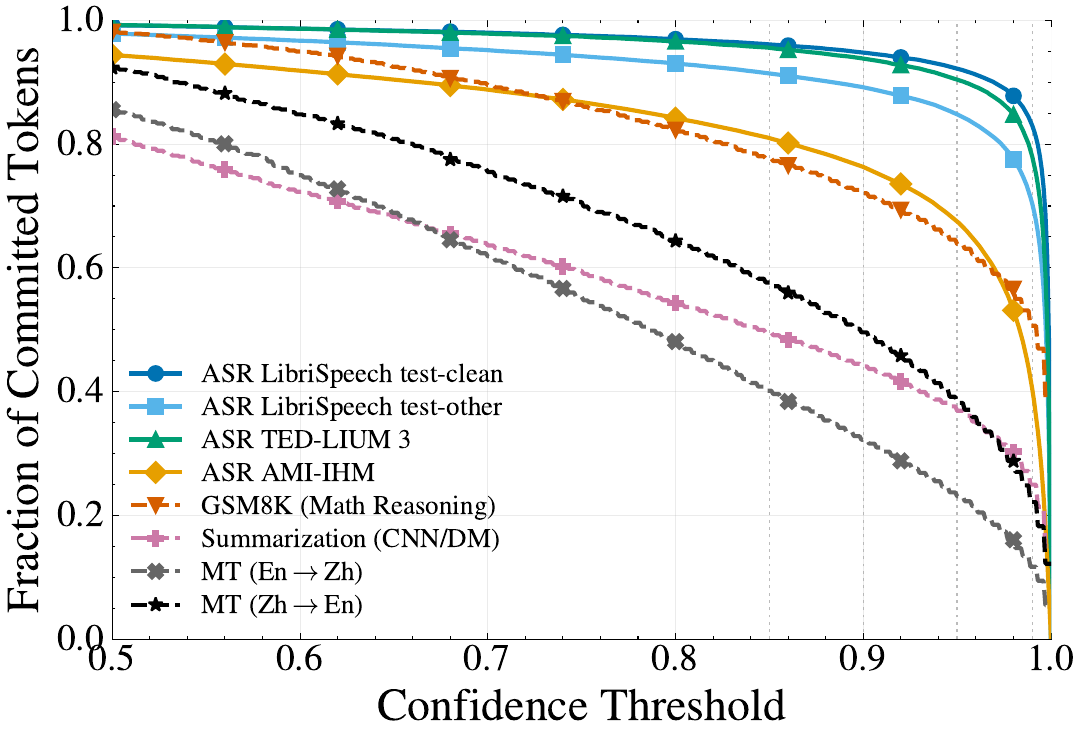}}
\vspace{-0.3cm}
\caption{Complementary cumulative distribution function (CCDF) of token confidence scores. The plot shows the fraction of tokens with a confidence score $S$ greater than or equal to a threshold $s$.}
\vspace{-0.5cm}
\label{fig:4}
\end{figure}

As shown in Fig.~\ref{fig:4}, confidence is generally more concentrated at high values in the evaluated ASR tasks than in the evaluated text tasks. At a confidence threshold of 0.90, 94.8\% and 89.2\% of tokens exceed the threshold on LibriSpeech test-clean and test-other, respectively, compared with 72.0\% on GSM8K~\cite{cobbe2021training} and 44.1\% on CNN/DailyMail~\cite{hermann2015teaching}, with lower fractions observed for WMT19 machine translation~\cite{barrault2019findings}. TED-LIUM 3 remains close to LibriSpeech, while AMI-IHM shows a larger decrease but still retains more high-confidence tokens than all evaluated text tasks at thresholds from 0.85 to 0.95.

This confidence concentration is consistent with the observed differences in decoding efficiency. Static thresholding can commit many high-confidence tokens at once, whereas the dynamic-threshold criterion becomes more restrictive as more tokens are selected: $c^{(k)} > 1 - f/(k+1)$ requires $c^{(k)}$ to approach one as $k$ increases. Fixed-number decoding, in contrast, is limited by its fixed per-round token budget.

\section{Conclusion}
We compared fixed-number, static-threshold, and dynamic-threshold decoding for DLM-based ASR across multiple English speech test sets and two DLM backbones. Across the evaluated settings, threshold-based decoding provides a better accuracy–speed trade-off than fixed-number decoding. Static and dynamic thresholding achieve comparable WER, with static thresholding generally yielding lower RTF. In the evaluated ASR settings, confidence scores are more concentrated at high values than in the examined text tasks, which is consistent with the observed early commitment behavior. The decoding-strategy trends are similar across the evaluated datasets and two DLM backbones, while confidence concentration decreases under domain shift. Future DLM decoding strategies for ASR may benefit from reducing late-stage refinement cost while maintaining recognition accuracy.

\section*{Limitations}
We do not evaluate accented speech, low-resource or multilingual ASR, or ASR models trained directly on spontaneous speech. Although we evaluate two DLM backbones, all systems are trained only on LibriSpeech 960h. Results on TED-LIUM 3 and AMI-IHM therefore reflect domain shift for LibriSpeech-trained models rather than performance after in-domain training on prepared talks or spontaneous meetings. Confidence concentration also weakens under domain shift and is lower on AMI-IHM than on the LibriSpeech test sets. Whether the observed pattern holds for models trained on spontaneous speech or under broader linguistic and acoustic conditions remains an open question.

\section*{Ethical Considerations}
This work uses publicly available datasets, pretrained models, and benchmarks, including LibriSpeech, TED-LIUM 3, AMI, WMT19, CNN/DailyMail, GSM8K, Whisper, LLaDA-8B-Instruct, Dream-7B, and LLaMA-3.1-8B. We use these resources solely for research purposes and follow their respective licenses, access conditions, and terms of use. We do not collect new human-subject data or conduct user studies. ASR systems may produce transcription errors and should be deployed with appropriate validation and human oversight, particularly in high-stakes settings.

\section*{Acknowledgments}
This work was supported in part by the National Research Foundation of Korea (NRF) grant funded by the Korea government (MSIT) (RS-2022-NR070162), and in part by the Institute of Information and Communications Technology Planning and Evaluation (IITP) grant funded by Korean Government (Ministry of Science and ICT) under Grant RS-2022-II220984.

\bibliography{custom}

\clearpage
\appendix

\begin{table}[t]
\centering
\small
\setlength{\tabcolsep}{4pt}
\renewcommand{\arraystretch}{1.08}
\begin{tabular}{@{}lp{4.8cm}@{}}
\toprule
\textbf{Setting} & \textbf{Configuration} \\
\midrule

\rowcolor{gray!12}
\multicolumn{2}{@{}l}{\textbf{Model}} \\
Speech encoder & Whisper-medium.en (frozen) \\
Adapter & Conv1d ($k=4$, $s=4$) + Linear $\rightarrow 4096$ \\
Decoder & LLaDA-8B-Instruct \\
LoRA & \{q, k, v, o\}, $r=16$, $\alpha=32$, dropout 0.05 \\
Answer canvas & 128 tokens; EOS padding; left-padded batches \\

\addlinespace[2pt]
\rowcolor{gray!12}
\multicolumn{2}{@{}l}{\textbf{Training}} \\
Objective & Cross-entropy \\
Optimizer & AdamW; lr $=10^{-4}$; $\beta_1=0.9$, $\beta_2=0.999$; wd $=0.01$ \\
Schedule & Cosine decay; 500 warm-up; 100K steps \\
Batching & $\leq$12 utterances/GPU; gradient accumulation 2 \\
Precision & bf16; DeepSpeed ZeRO-2; gradient clipping 1.0 \\

\addlinespace[2pt]
\rowcolor{gray!12}
\multicolumn{2}{@{}l}{\textbf{Data and Compute}} \\
Training data & LibriSpeech 960h \\
Hardware & 8$\times$ NVIDIA RTX A6000 \\
Training time & $\approx$25.8 h \\
Seed & 10086 \\

\bottomrule
\end{tabular}
\caption{Training configuration of the ASR system.}
\label{tab:a1}
\end{table}
\begin{table}[t]
\centering
\scriptsize
\setlength{\tabcolsep}{1.1pt}
\renewcommand{\arraystretch}{1.05}
\begin{tabular}{@{}cccccccc@{}}
\toprule
& \multicolumn{7}{c}{$k$} \\
\cmidrule(lr){2-8}
$B$ & 1 & 2 & 4 & 8 & 16 & 32 & 128 \\
\midrule
4   & 2.79/.205 & 3.62/.113 & 4.80/.067 & --        & --         & --          & -- \\
8   & 2.93/.209 & 3.83/.111 & 4.47/.062 & 6.70/.038 & --         & --          & -- \\
16  & 3.01/.242 & 3.47/.122 & 4.02/.068 & 5.68/.038 & 9.07/.024  & --          & -- \\
32  & 3.05/.298 & 3.44/.154 & 3.91/.080 & 5.01/.046 & 7.35/.028  & 12.10/.019  & -- \\
64  & 3.32/.489 & 3.57/.228 & 4.03/.114 & 4.80/.064 & 6.50/.037  & 10.50/.022  & -- \\
128 & 4.23/.627 & 4.23/.301 & 4.75/.148 & 5.42/.084 & 6.64/.052  & 9.92/.033   & 14.50/.014 \\
\bottomrule
\end{tabular}
\caption{Fixed-number decoding on LibriSpeech test-clean.
Each cell reports WER (\%) / RTF. Dashes denote inapplicable settings.}
\label{tab:a2}
\end{table}

\section{Experimental Details}
\label{app:details}

\subsection{Model and training.}
Table~\ref{tab:a1} provides the full training configuration. Unless otherwise stated, all reported results are obtained from a single run with seed 10086. The Dream-7B variant in Section~\ref{sec:backbone} replaces only the decoder, with all other settings unchanged.

\subsection{Datasets.}
All models are trained on LibriSpeech 960h. Evaluation is conducted on LibriSpeech test-clean (2,620 utterances, 5.4 h), test-other (2,939 utterances, 5.3 h), TED-LIUM 3 test (1,154 utterances, 2.6 h), and AMI-IHM test (12,643 segments, 8.7 h). TED-LIUM 3 and AMI-IHM are evaluated using the LibriSpeech-trained model without adaptation. 

\subsection{Inference and latency measurement.}
Decoding uses the dual prefix--suffix KV-cache of Fast-dLLM with a fixed 128-token answer canvas. Once an EOS token is committed, all remaining positions are filled with EOS and subsequent blocks are skipped. All latency measurements are performed on a single idle NVIDIA RTX A6000 with batch size 1 and one evaluation job per GPU. RTF is computed as the total inference time, including the speech encoder, divided by the total audio duration. Average NFE is the mean number of decoder forward passes per utterance.

\subsection{AR baselines.}
AR (LLaDA-8B) decodes the same fine-tuned DLM one token at a time from left to right and serves as the reliability reference. AR (LLaMA-3.1-8B) is a LoRA-tuned autoregressive decoder trained with the same speech encoder, adapter, training data, and optimization schedule.

\begin{figure*}[t]
\centering
\centerline{\includegraphics[width=14cm]{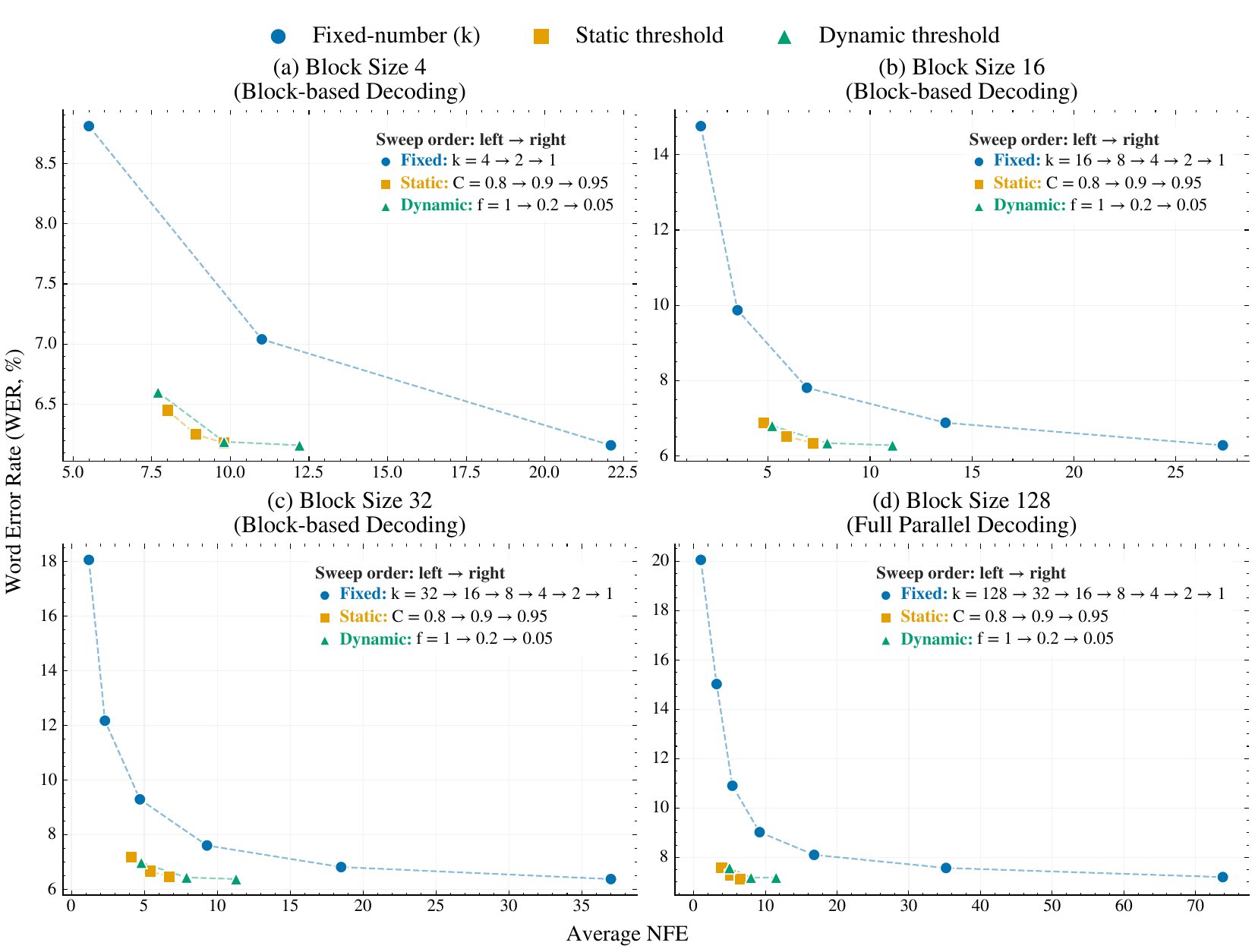}}
\vspace{-0.3cm}
\caption{WER–NFE trade-off on LibriSpeech test-other across decoding strategies and block sizes. Parameters in each panel are listed in order of increasing NFE. Lower WER and NFE are preferred.}
\label{fig:a1}
\vspace{-0.5cm}
\end{figure*}

\begin{table}[t]
\centering
\small
\setlength{\tabcolsep}{3.2pt}
\renewcommand{\arraystretch}{1.05}
\begin{tabular}{@{}cccc@{}}
\toprule
& \multicolumn{3}{c}{$C$} \\
\cmidrule(lr){2-4}
$B$ & 0.8 & 0.9 & 0.95 \\
\midrule
4   & 3.17/.081 & 2.85/.086 & 2.81/.100 \\
8   & 3.53/.053 & 3.07/.058 & 2.96/.066 \\
16  & 3.72/.039 & 3.15/.047 & 3.04/.052 \\
32  & 3.95/.031 & 3.25/.038 & 3.11/.045 \\
64  & 4.37/.029 & 3.62/.035 & 3.50/.043 \\
128 & 4.93/.029 & 4.28/.038 & 4.13/.046 \\
\bottomrule
\end{tabular}
\caption{Static thresholding on LibriSpeech test-clean.
Each cell reports WER (\%) / RTF.}
\label{tab:a3}
\end{table}
\begin{table}[t]
\centering
\small
\setlength{\tabcolsep}{3.2pt}
\renewcommand{\arraystretch}{1.05}
\begin{tabular}{@{}cccc@{}}
\toprule
& \multicolumn{3}{c}{$f$} \\
\cmidrule(lr){2-4}
$B$ & 1.0 & 0.2 & 0.05 \\
\midrule
4   & 3.32/.081 & 2.81/.093 & 2.79/.110 \\
8   & 3.52/.059 & 2.97/.075 & 2.93/.096 \\
16  & 3.41/.044 & 3.03/.061 & 3.02/.083 \\
32  & 3.51/.039 & 3.12/.054 & 3.05/.081 \\
64  & 3.88/.040 & 3.44/.061 & 3.31/.082 \\
128 & 4.47/.042 & 4.18/.063 & 4.23/.089 \\
\bottomrule
\end{tabular}
\caption{Dynamic thresholding on LibriSpeech test-clean.
Each cell reports WER (\%) / RTF.}
\label{tab:a4}
\end{table}
\begin{table}[t]
\centering
\scriptsize
\setlength{\tabcolsep}{1.2pt}
\renewcommand{\arraystretch}{1.05}
\begin{tabular}{@{}cccccccc@{}}
\toprule
& \multicolumn{7}{c}{$k$} \\
\cmidrule(lr){2-8}
$B$ & 1 & 2 & 4 & 8 & 16 & 32 & 128 \\
\midrule
4
& 6.16/22.1
& 7.04/11.0
& 8.81/5.5
& --
& --
& --
& -- \\

16
& 6.28/27.3
& 6.88/13.7
& 7.81/6.9
& 9.87/3.5
& 14.76/1.7
& --
& -- \\

32
& 6.37/37.0
& 6.81/18.5
& 7.60/9.3
& 9.29/4.7
& 12.17/2.3
& 18.06/1.2
& -- \\

128
& 7.20/73.8
& 7.57/35.2
& 8.10/16.8
& 9.02/9.2
& 10.90/5.4
& 15.02/3.2
& 20.05/1.0 \\
\bottomrule
\end{tabular}
\caption{Fixed-number decoding on LibriSpeech test-other.
Each cell reports WER (\%) / average NFE. Dashes denote inapplicable settings.}
\label{tab:a5}
\end{table}
\begin{table}[t]
\centering
\small
\setlength{\tabcolsep}{4pt}
\renewcommand{\arraystretch}{1.05}
\begin{tabular}{@{}cccc@{}}
\toprule
& \multicolumn{3}{c}{$C$} \\
\cmidrule(lr){2-4}
$B$ & 0.8 & 0.9 & 0.95 \\
\midrule
4   & 6.45/8.0 & 6.25/8.9 & 6.18/9.8 \\
16  & 6.87/4.8 & 6.52/5.9 & 6.34/7.2 \\
32  & 7.17/4.1 & 6.66/5.4 & 6.45/6.7 \\
128 & 7.57/3.9 & 7.28/5.1 & 7.12/6.5 \\
\bottomrule
\end{tabular}
\caption{Static thresholding on LibriSpeech test-other.
Each cell reports WER (\%) / average NFE.}
\label{tab:a6}
\end{table}
\begin{table}[t]
\centering
\small
\setlength{\tabcolsep}{4pt}
\renewcommand{\arraystretch}{1.05}
\begin{tabular}{@{}cccc@{}}
\toprule
& \multicolumn{3}{c}{$f$} \\
\cmidrule(lr){2-4}
$B$ & 1.0 & 0.2 & 0.05 \\
\midrule
4   & 6.60/7.7 & 6.19/9.8 & 6.16/12.2 \\
16  & 6.80/5.2 & 6.34/7.9 & 6.28/11.1 \\
32  & 6.97/4.8 & 6.43/7.9 & 6.37/11.3 \\
128 & 7.57/5.0 & 7.18/8.0 & 7.18/11.5 \\
\bottomrule
\end{tabular}
\caption{Dynamic thresholding on LibriSpeech test-other.
Each cell reports WER (\%) / average NFE.}
\label{tab:a7}
\end{table}
\begin{table}[t]
\centering
\small
\setlength{\tabcolsep}{4pt}
\renewcommand{\arraystretch}{1.05}
\begin{tabular}{@{}llccc@{}}
\toprule
Test set & Backbone & Static & Dynamic & Fixed-$k$ \\
\midrule

\rowcolor{gray!12}
\multicolumn{5}{@{}l}{\textit{Table~\ref{tab:1}}} \\
test-other & -- & 7.2 & 7.9 & 13.7 \\
TED-LIUM 3 & -- & 7.0 & 8.0 & 16.8 \\
AMI-IHM    & -- & 6.7 & 7.1 & 9.2 \\

\addlinespace[2pt]
\rowcolor{gray!12}
\multicolumn{5}{@{}l}{\textit{Table~\ref{tab:2}}} \\
test-clean & LLaDA-8B & 5.0 & 6.2 & 9.8 \\
           & Dream-7B & 7.0 & 8.1 & 10.8 \\
test-other & LLaDA-8B & 6.7 & 7.9 & 9.3 \\
           & Dream-7B & 8.7 & 9.6 & 10.4 \\
\bottomrule
\end{tabular}
\caption{Average NFE per utterance for the main-text configurations in
Tables~\ref{tab:1} and~\ref{tab:2}.}
\label{tab:a8}
\end{table}
\begin{table}[t]
\centering
\small
\setlength{\tabcolsep}{3pt}
\begin{tabular}{lcccc}
\toprule
\multirow{2}{*}{Model} & \multicolumn{2}{c}{WER (\%)} & \multicolumn{2}{c}{RTF} \\
\cmidrule(lr){2-3}\cmidrule(lr){4-5}
 & clean & other & clean & other \\
\midrule
Whisper-medium.en (greedy)  & 3.1$^\dagger$ & 6.3$^\dagger$ & 0.078 & 0.079 \\
Whisper-medium.en (beam 5)  & 3.0$^\dagger$ & 5.7$^\dagger$ & 0.085 & 0.087 \\
Whisper-large-v3 (greedy)   & 2.17 & 4.05 & 0.087 & 0.103 \\
Whisper-large-v3 (beam 5)   & 2.00 & 3.71 & 0.125 & 0.128 \\
\midrule
Ours (static $C$=0.95, $B$=4)  & 2.81 & 6.18 & 0.100 & 0.107 \\
Ours (static $C$=0.95, $B$=16) & 3.04 & 6.34 & 0.052 & 0.072 \\
\bottomrule
\end{tabular}
\caption{Comparison with Whisper baselines on LibriSpeech. WERs marked with
$\dagger$ are taken from \citet{radford2023robust}. All other WERs and all
RTF values were measured by us on a single idle NVIDIA RTX A6000 with batch
size 1. Our block-16 configuration reaches WER close to Whisper-medium.en
greedy decoding at a lower RTF, while Whisper-large-v3 attains lower WER at
a higher RTF.}
\label{tab:a9}
\end{table}
\begin{table}[t]
\centering
\small
\setlength{\tabcolsep}{3pt}
\begin{tabular}{lrcccc}
\toprule
Task & $N$ & $\geq$0.85 & $\geq$0.90 & $\geq$0.95 & $\geq$0.99 \\
\midrule
test-clean & 50,046 & 96.1 & 94.8 & 92.2 & 83.0 \\
test-other & 51,514 & 91.4 & 89.2 & 84.8 & 70.4 \\
TED-LIUM 3 & 25,195 & 95.6 & 93.8 & 90.4 & 78.0 \\
AMI-IHM    & 83,000 & 81.0 & 76.3 & 67.5 & 40.4 \\
\midrule
GSM8K      & 16,000 & 77.7 & 72.0 & 63.9 & 50.7 \\
CNN/DailyMail & 15,999 & 49.6 & 44.1 & 36.9 & 25.0 \\
MT (En$\rightarrow$Zh) & 9,667 & 40.2 & 32.2 & 23.1 & 11.7 \\
MT (Zh$\rightarrow$En) & 10,854 & 57.6 & 49.6 & 38.2 & 22.2 \\
\bottomrule
\end{tabular}
\caption{Fraction (\%) of committed tokens with confidence at or above each
threshold under sequential confidence-ordered decoding (block size 32,
first 32-token block preceding the first EOS, EOS excluded). ASR tasks are
listed above the rule and text tasks below it.}
\label{tab:a10}
\end{table}

\section{Additional Results and Analyses}
\label{app:results}

\subsection{Full Accuracy--Speed Sweep on LibriSpeech test-clean}
\label{app:sweep}

Tables~\ref{tab:a2}--\ref{tab:a4} report the numerical values underlying Fig.~\ref{fig:1}. Each table covers one commitment strategy over its full hyperparameter range and all six block sizes, with WER (\%) and RTF reported for each configuration. The AR reference points are independent of block size: AR (LLaDA-8B) achieves 2.94\% WER at RTF 0.276, while AR (LLaMA-3.1-8B) achieves 2.78\% WER at RTF 0.160. Across block sizes up to 64, stricter commitment criteria consistently yield lower WER within each strategy. 

\subsection{Full Accuracy--NFE Sweep on LibriSpeech test-other}
\label{app:sweepother}

Tables~\ref{tab:a5}--\ref{tab:a7} report the corresponding
sweep on LibriSpeech test-other for block sizes $B\in\{4,16,32,128\}$, and Fig.~\ref{fig:a1} shows the resulting trade-off curves. Each cell reports WER (\%) / average NFE, using NFE as a hardware-independent measure of decoding cost. RTF values for representative test-other configurations are reported in Table~1. The overall ordering observed on test-clean is preserved: threshold-based decoding consistently provides lower WER than fixed-number decoding at comparable decoding cost, while static and dynamic thresholding remain closely matched.

\subsection{Average NFE for Main-Text Configurations}
\label{app:nfe}

Tables~\ref{tab:1} and~\ref{tab:2} report RTF as the primary cost metric because it reflects wall-clock latency on the target hardware. For completeness, Table~\ref{tab:a8} reports the average number of decoder forward passes (NFE) per utterance for the same configurations. NFE follows the same overall efficiency trend as RTF: static thresholding uses the fewest forward passes in every setting, while dynamic thresholding requires 0.4--1.2 additional passes per utterance at nearly matched WER. Fixed-number decoding requires more forward passes than static thresholding while also yielding higher WER.

\subsection{Comparison with Whisper Baselines}
\label{app:whisper}

Table~\ref{tab:a9} compares our system with Whisper-medium.en and Whisper-large-v3. All RTF values are measured under the same latency protocol. On LibriSpeech test-clean, our block-16 static-threshold configuration achieves WER close to Whisper-medium.en greedy decoding at lower RTF, while Whisper-large-v3 achieves lower WER at higher RTF. This comparison is intended to contextualize the absolute performance rather than claim state-of-the-art accuracy, as our system uses the Whisper-medium.en encoder and is trained only on LibriSpeech 960h.

\subsection{Confidence Statistics Across ASR and Text Tasks}
\label{app:confidence}

Table~\ref{tab:a10} reports the statistics underlying Fig.~\ref{fig:4}. Following Section~\ref{sec:confidence_dist}, we use sequential confidence-ordered decoding with block size 32, committing only the highest-confidence token at each step. Commit-time confidence is recorded for positions in the first 32-token block preceding the first EOS token, with EOS excluded. The text tasks use the base LLaDA-8B-Instruct model with its chat template on 500 examples each from GSM8K, CNN/DailyMail, and WMT19 in both translation directions. The ASR tasks use the LibriSpeech-trained model on the full test sets. $N$ denotes the number of pooled token commitments per task.

Across thresholds from 0.85 to 0.95, every evaluated ASR dataset retains a larger fraction of high-confidence tokens than any of the text tasks. AMI-IHM shows the largest decrease among the ASR datasets under domain shift and falls below GSM8K at the 0.99 threshold, indicating that confidence concentration at the extreme upper end is more sensitive to acoustic conditions.

\subsection{Statistical Validation}
\label{app:stats}

\paragraph{Utterance-level NLL--WER association.}
To statistically assess the relationship between cumulative uncertainty and recognition error discussed in Fig.~\ref{fig:2}, we compute the Spearman rank correlation between the final cumulative committed-token NLL and utterance-level WER under dynamic thresholding ($f=1.0$, block size 32). The correlation is $\rho=0.627$ on LibriSpeech test-clean ($n=2{,}620$) and $\rho=0.672$ on test-other ($n=2{,}939$). These results support a positive utterance-level association between cumulative uncertainty and recognition error: utterances with larger accumulated uncertainty tend to have higher WER.

Because cumulative committed-token NLL reflects the model's confidence in its own predictions rather than prediction correctness, it need not track WER uniformly across decoding configurations. This limitation can become relevant in near-single-shot fully parallel settings, where many tokens are committed from a largely masked context. We therefore use the per-utterance correlations above as the primary statistical evidence supporting the uncertainty-trajectory analysis, rather than drawing conclusions from an across-configuration correlation.

\paragraph{Paired WER comparisons on LibriSpeech test-other.}
We further assess the WER differences between static-threshold and fixed-number decoding using a paired utterance-level bootstrap with 10,000 resamples on LibriSpeech test-other with block size 16. For the configurations reported in Table~\ref{tab:1}, static thresholding ($C=0.95$) reduces WER relative to fixed-number decoding with $k=2$ by 0.54 absolute percentage points, with a 95\% confidence interval of [0.44, 0.64] and $p<10^{-4}$. The average NFE is 7.2 for static thresholding and 13.7 for fixed-number decoding.

We additionally compare static thresholding against fixed-number decoding with $k=4$, which approximately matches its decoding computation (7.2 versus 6.9 average NFE). At this matched-computation operating point, static thresholding reduces WER by 1.47 absolute percentage points, with a 95\% confidence interval of [1.25, 1.65] and $p<10^{-4}$. These results indicate that the WER differences between static thresholding and the tested fixed-number configurations are statistically significant both for the main-text configuration and under approximately matched decoding computation. We restrict claims of statistical significance to these two tested comparisons.


\end{document}